\documentclass{ab2018}
\usepackage{url}
\usepackage{graphicx}
\usepackage{natbib}

\begin{document} 

\title{Internal Kinematics of the Seyfert Galaxy  Mkn\,938}

\author{V.~L.Afanasiev, A.V.~Moiseev and A.~A.~Smirnova}

\institute{Special Astrophysical Observatory, Russian Academy of Sciences, Nizhnij Arkhyz, 369167  Russia}

\titlerunning{Internal Kinematics of   Mkn\,938}

\authorrunning{Afanasiev et al.}

\date{September 24, 2019/Revised: November 14, 2019}
\offprints{Victor Afanasiev  \email{vafan@sao.ru} }

\abstract{
We present  the results of a detailed study of the central part of
the Seyfert galaxy Mkn\,938. Observational data were obtained with
the 6-m telescope of the Special Astrophysical Observatory of the
Russian Academy of Sciences using integral-field spectrograph MPFS
and a scanning Fabry--Perot interferometer. Mkn\,938 is
interesting for being a result of a merger of two gas-rich
galaxies, and we observe the final stage of this interaction 
accompanied with an extremely powerful burst of star formation and
nuclear activity. Our analysis of the kinematics of gas and stars
revealed the presence of gas outflow in the circumnuclear region
Mkn\,938 with  velocities ranging from $-370$ to
$-480$~km\,s$^{-1}$, and allowed us for the first time to map the
high-velocity galactic wind in Na\,D absorption line on large
spatial scale in this galaxy.
}

\maketitle

\section{INTRODUCTION}
\label{S:int}

Studies of merging gas-rich galaxies provide an insight into  the
role that the dissipative processes along with gravitational
processes play in the evolution of galaxies and galaxy
systems~\citep{Sch07}. In  recent decades great progress has been
achieved in the  understanding of the general role of mergers in
the formation of early-type galaxies, and it is now generally
believed that many of them are products of mergers of gas-rich
galaxies. It is, however, not entirely  clear how a has-rich
system eventually gets rid of gas and how this process is affected
by a burst of star formation at the center and the presence of
active nucleus. Bright  star-forming regions and outflows form
various components along the line-of-sight, which cannot be
resolved in the sky plane. The application of the methods of
two-dimensional (often referred  to as 3D or integral-field)
spectroscopy makes it possible to study such objects in detail. In
this paper we report the results of 3D spectroscopy
of the central  part of the Mkn\,938 galaxy. The main parameters
of Mkn\,938 are summarized in Table~\ref{param}.

\begin{table}[]
\centering
\caption[]{The summary of Mkn\,938} \label{param}
\begin{tabular}{l|r}
\hline
Activity type & Sy\,2 \\
Morphological type (NED) & Pec \\
Systemic velocity~\citep{Rot06} & $5881$~km\,s$^{-1}$  \\
Adopted distance  & $85.2$~Mpc \\
Image scale  & $395$~pc/\arcsec \\
 $M_V$~\citep{Sch07}  & $-21.42$ \\
HI mass~\citep{Kand}  & $5.3\times10^9~M_{\odot}$\\
Maximum rotation velocity  &   \\
 (our measurements for  $i=38\degr$)  & $225\pm6$~km\,s$^{-1}$  \\
\hline
\end{tabular}
\end{table}

We classified the interacting galaxy Mkn\,938 \linebreak
(=\,NGC\,34\,=\,VV\,850), which has a peculiar morpho\-lo\-gy and
active nucleus, as Sy\,2~\citep{afan80}. This classification
was later confirmed  \citep{Dah85,ver86}. On the contrary,
~\citet{ost83} classified it as emission-line galaxy dominated by
star-forming regions, and~\citet{mul96} observed the galaxy in
narrow-band filters and found that in Mkn\,938
[O\,III]\,$\lambda$5007\ emission is weak compared to most of the
known Seyfert galaxies. In addition, the above authors found that
the bright H$\alpha$ emission shows up throughout the entire
galaxy. This means that ionization of most of the gas is not
associated with any Seyfert activity. Morphologically, the galaxy
appears to undergo a merger, as is evidenced by the presence of
tidal tails. This conclusion is corroborated by 8.8- and
12.5-$\mu$m infrared images of Mkn\,938~\citep{mil96}, where a
binary source with a separation of~$1\farcs2$ was found in the
nuclear region of the galaxy. Preliminary results of the
integral-field spectroscopy of the central part of this galaxy
carried out with the 6-m telescope of the Special Astrophysical
Observatory of the Russian Academy of Sciences~\citep{raf00}
revealed signs of both star formation and gas outflows. In
particular, two dynamic centers were found in the  Na\,D line with
a separation of $1\arcsec$, which coincide with the double source
found in the infrared. The extended H\,I structure found in the
VLA image of this galaxy~\citep{Fer10} and the circumnuclear disk
found in CO observations performed with ALMA~\citep{Xu14} suggest
that the observed galaxy is a gas-rich interacting system. Deep
optical images~\citep{Sch07} also suggest that Mkn\,938 is likely
a product of a merger of galaxies with a mass ratio of 1$\colon$3.
The high IR luminosity  indicates that star formation
dominates at the center and with a minor contribution from
AGN~\citep{Gon99}. This conclusion is confirmed by a detailed
study of the infrared spectrum of Mkn\,938~\citep{Esq12}, which
estimates the bolometric contribution of AGN to the total IR
luminosity to be~2\%. Furthermore, IR images show that star
formation occurs within the central \mbox{0.5--2~kpc}, where
merger traces can be seen. However, these authors believe that the
presence of an AGN is necessary for explaining the hard X-ray
luminosity.

\section{OBSERVATIONS}

We observed Mkn\,938 in the optical with the 6\mbox{-}m telescope
of the Special Astrophysical Observatory of the Russian Academy of
Sciences (SAO RAS) within the framework of a program of spectrophotometry of
Seyfert galaxies. The heliocentric velocity of the center of the
galaxy determined from absorption-line spectra~\citep{Rot06} is
\mbox{$5881\pm2$~km\,s$^{-1}$}, which implies a distance of
$85.2$~Mpc for \mbox{$H_0 = 73$~km\,s$^{-1}\,$Mpc$^{-1}$}. This
corresponds to a sky-plane scale of $395$~pc/\arcsec.

Table~\ref{log} presents the log of observations of  Mkn\,938 at the 6-m telescope:   date
of observations, total exposure, size of the resulting data cube,
seeing, the instrument employed, disperser or filter, spectral
range and spectral resolution. A description of the instruments
employed  (MPFS integral-field spectrograph and focal reducer with
a Fabry--Perot interferometer -- FPI)  can be found
in~\citet{Af01}. The data acquired with both instruments was
reduced using the technique described in our earlier
papers~\citet{Smirnova2006,Smirnova2010}.

\begin{table*}[]
\captionstyle{normal}
\caption[]{Log of observations}\label{log}
\medskip
\begin{tabular}{c|c|c|c|c|c|c|c}
\hline
Date of&Total exp,&Size of&Seeing,& Device &Disperser&Spectral&Spectral \\
observation&sec.&data cube&arcsec&  &or Filter &coverage, \AA&resolution, \AA\\
\hline
1999/12/04& 2400&512$\times$512$\times$12   &   2& focal reducer &FP260    &6675--7005& 2.5\\
2009/10/18& 9600&16$\times$16$\times$3000   &   1.5& MPFS    &1200/17.7&4790--7740& 3\\
\hline
\end{tabular}
\end{table*}

\subsection{Fabry--Perot Interferometer}

As a result of observations with a scanning FPI a total of 12
interferograms were acquired that fill sequentially the free
spectral interval  ($\Delta\lambda\approx29~$\AA{} for this
interference order). The spectral domain in the neighborhood of
redshifted  H$\alpha$ line was selected using a narrow-band
filter. To reduce readout time and increase the signal-to-noise
ratio, TK~1025 1K$\times$1K CCD was read out in the 2$\times$2
instrumental-binning mode. The size of the field of view was
5\farcm8 with a pixel scale of 0\farcs68 pixel$^{-1}$.

\begin{figure*}[]
\centerline{\includegraphics[width=15cm]{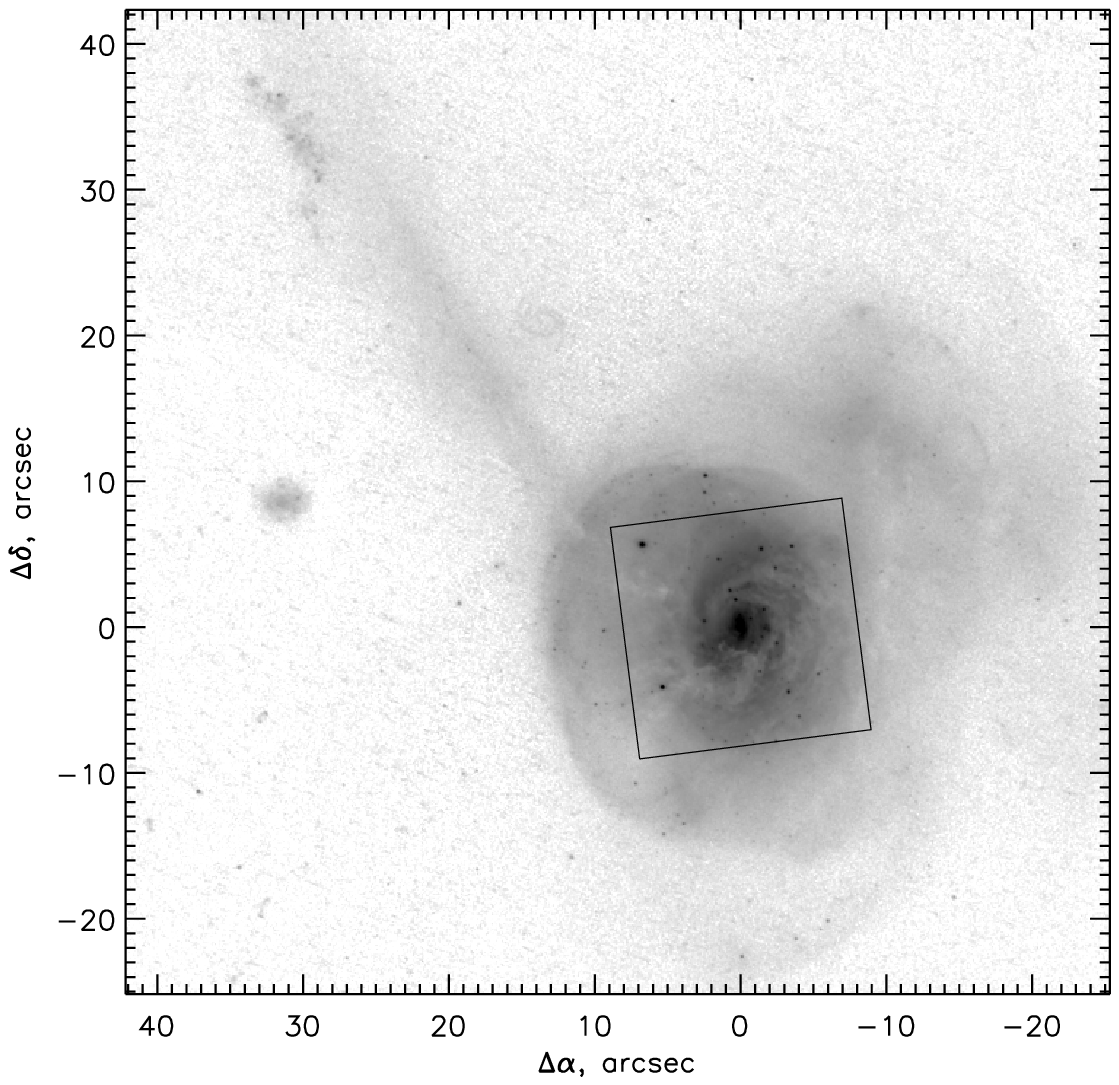}}
\centerline{
\includegraphics[height=8.5cm]{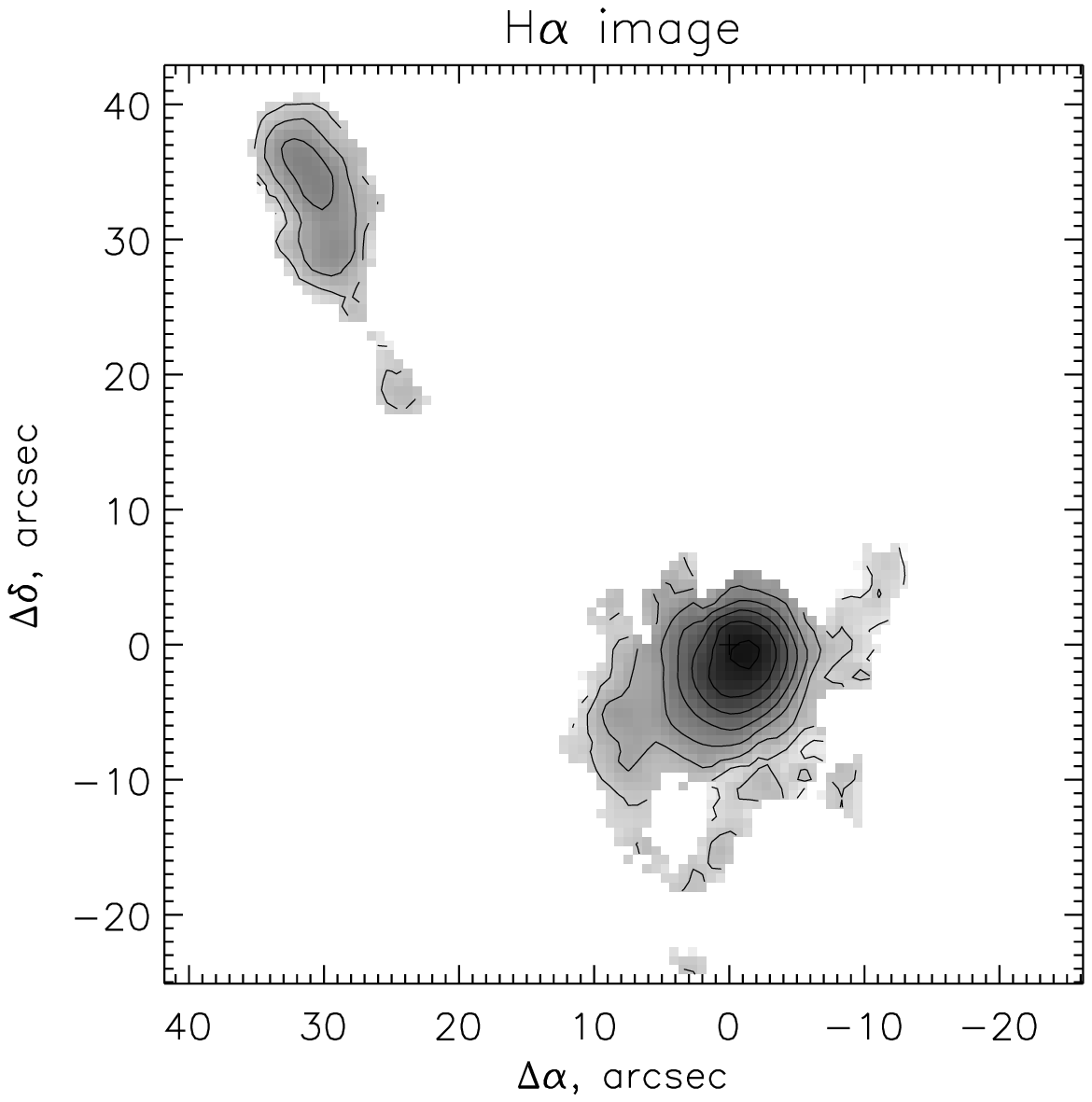}
\includegraphics[height=8.5cm]{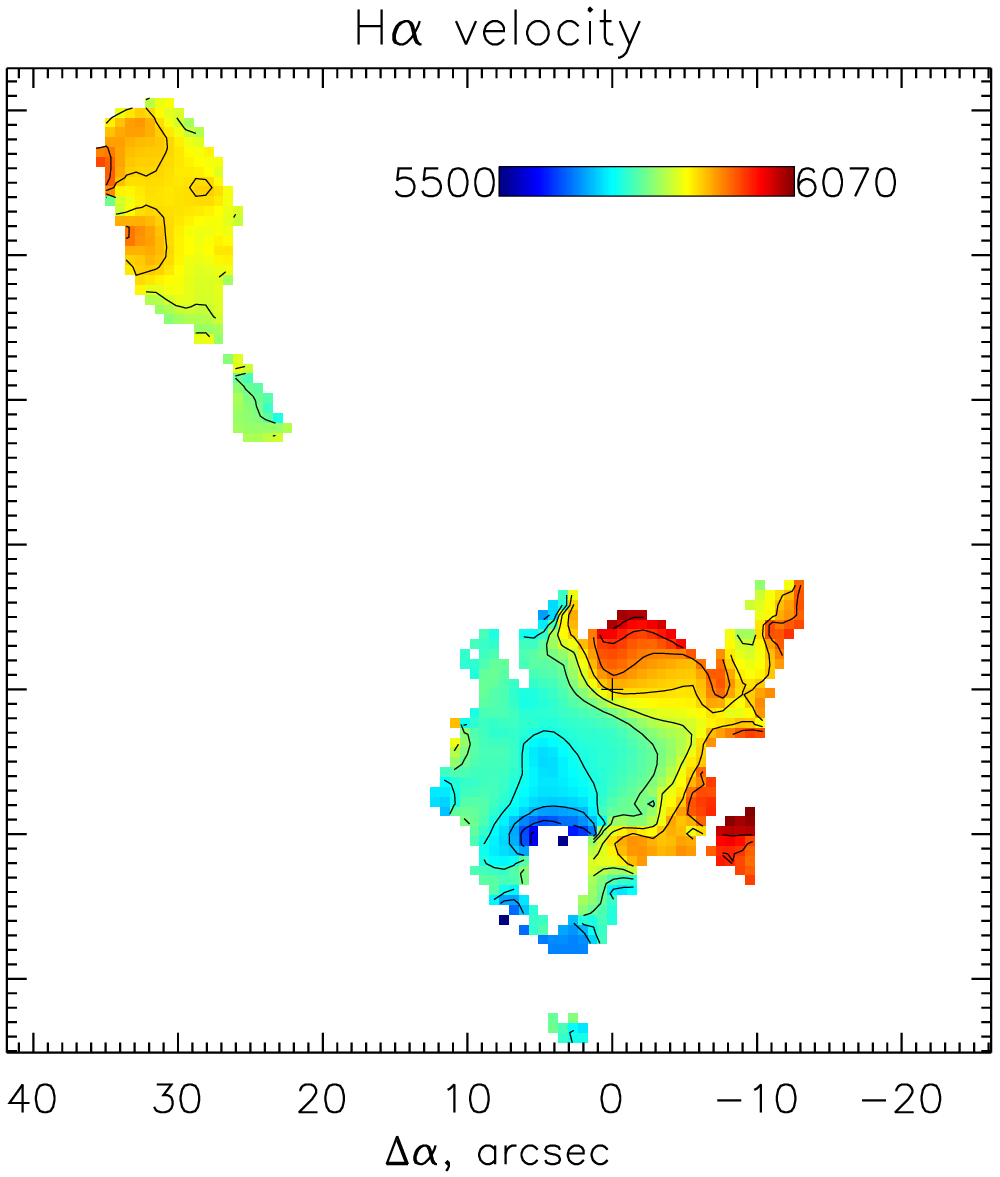}
}
\caption{F435W-band HST image of  Mkn\,938 obtained with ASC
camera (the top panel). The square indicates the field observed with the MPFS. The bottom panels show
results of observations made with the scanning FPI: the
H$\alpha$\ brightness distribution and the velocity field in this
line, respectively. The velocity scale is in km\,s$^{-1}$. The
cross indicates the center of continuum isophotes near H$\alpha$.}
\label{fig:IFP}
\end{figure*}

As a result of data reduction we obtained  data cube where each
pixel in the field of view contains a  12-channel spectrum. We
used software Aladin~\citep{Aladin2014} for astrometric calibration.
We fitted emission lines by Gaussian profiles to construct the
brightness distributions in the H$\alpha$\ line and the
line-of-sight velocity field of ionized gas shown in Fig.~\ref{fig:IFP}.
For technical reasons we had to limit our observations to a few
interferograms providing a spectrum sampling close to the
width of the instrumental contour, and therefore we could not map
the distribution of velocity dispersion in the emission line.

The emission line is detected inside the bright stellar disk of
the galaxy ($r<15\arcsec$--$20\arcsec$), and in the emission-line
``island'' located at a projected distance of
$30\arcsec$--$55\arcsec$ Northeast of the nucleus. This ``island''
is  quite conspicuous on broadband Hubble Space Telescope images
acquired with the ACS camera and published in~\citet{Kim2013}, and
also on earlier images from~\citet{Sch07}. Fig.~\ref{fig:IFP}  shows an image taken in the  F438W blue filter. This
emission feature must be a region of intense star formation
associated with an extended tidal tail. This is possibly a tidal
galaxy with a size of about~6~kpc.

\subsection{MPFS integral-field spectrograph}
MPFS simultaneously records spectra from 256 spatial elements
formed by a  16\arcsec$\times$16\arcsec lens square array placed
in the focal plane. In our observations each lens had a size of
1\arcsec$\times$1\arcsec. This lens array forms an array of
micropupils. Optical fibers reform these micropupils into a
pseudoslit at the spectrograph entrance. The sky background
spectra are acquired with other fibers located  4\arcmin\ from the
lens array. The   detector was a 2K$\times$4.5K CCD E2V\,42-90.
Fig.~\ref{fig:IFP} shows the location of the MPFS array on the
galaxy image.

After primary reduction the results had the form of data cubes
with a  16$\times$16  element field of view where each element
contained a spectrum with bright emission lines typical of Seyfert
galaxies---H$\beta$, [O\,III]\,$\lambda$4959,\,5007, H$\alpha$,
[N\,II]\,$\lambda$6548,\,6584 and  [S\,II]\,$\lambda$6717,\,6731,
and absorption lines H$\beta$, Mg\,I and Na\,D.

To measure the fluxes and radial velocities, we fitted emission
lines by Gaussians. A single-component model describes the center
of the galaxy far from the line center quite well, however, in the
central region ($r<4\arcsec$, see Section~\ref{sec:outflow})
a second broad component shows up with an amplitude amounting to
$\sim1/4$ of that of the main component. Hereafter we show the velocity
fields and flux maps for the main (narrow) component. To take into
account the contribution of high-contrast absorptions due to
Balmer lines of the stellar population, we fitted the H$\beta$\
absorption by another Gaussian. We assumed  the velocity and width
of the absorption feature in the region of H$\alpha+$[N\,II] lines
to be the same as in the  H$\beta$ neighborhood, and chose the
amplitude so as to preserve the same equivalent width of the Balmer
lines.

\begin{figure}[t]
\includegraphics[width=0.85\columnwidth]{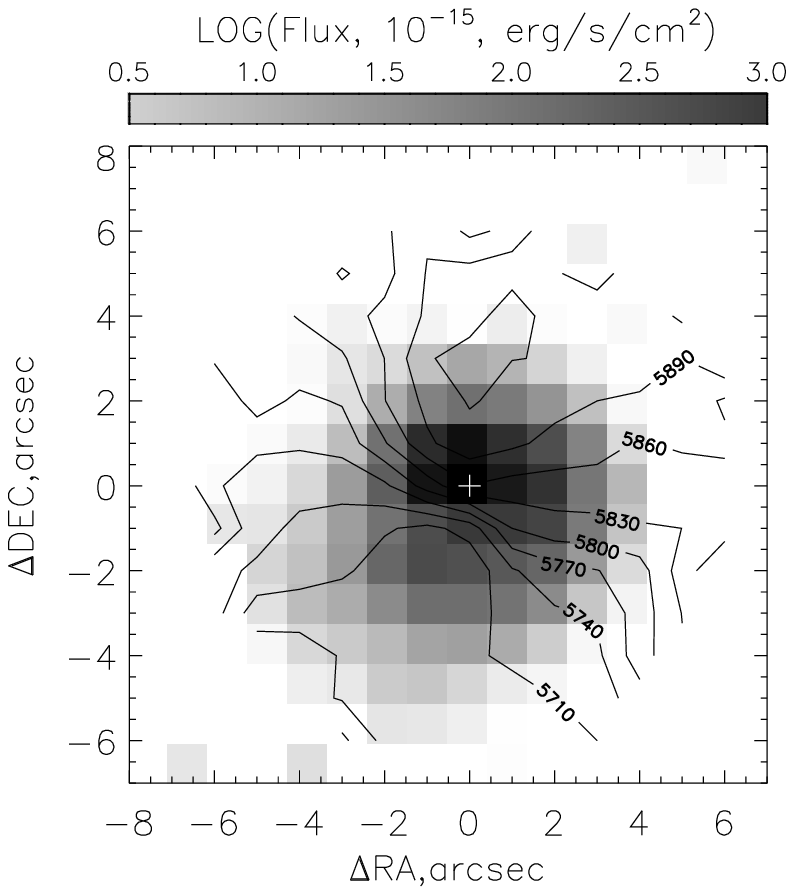}
\caption{H$\alpha$ velocity field of  Mkn\,938 superimposed on the
$V$-band continuum image of the central part of the galaxy.}
\label{fig:Halpha}
\end{figure}

\begin{figure}[t]
\centering
\includegraphics[width=0.9\columnwidth]{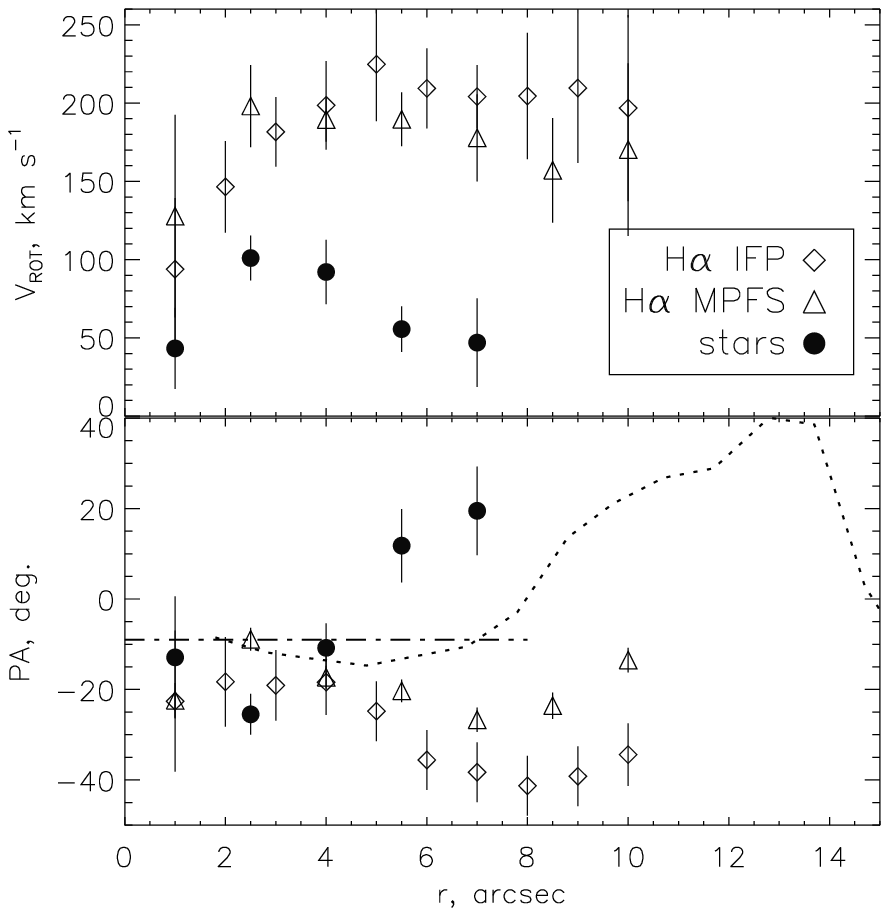}
\caption{Results of an analysis of the velocity fields of gas and
stars: the rotation curves (the top panel) and the position angle
of the kinematic major axis  (the bottom panel). The dashed line
shows the change of orientation of the major axis of the $K$-band
isopthotes~\citep{Rot04}. The dashed-and-dotted line shows the
orientation of the major axis of the inner disk according
to~\citet{Sch07}. } \label{fig:rc}
\end{figure}

We constructed the velocity  field of the stellar population by
applying the cross-correlation technique~\citep{Moiseev2001} using
standard spectra of stars in the  5015--5400~\AA, wavelength
interval, which contains high-contrast  Mg\,I and Fe\,I absorption
features. We used the same method to analyze the line-of-sight velocity
profiles in the absorption line of the  Na\,D, because these
radial velocities differ widely from the radial velocities
observed both in the stellar population and in ionized gas (see
Section~\ref{sec:NaD} below). We used night-sky emission lines
recorded in the field of view of MPFS as the template Na\,D
spectrum.

\section{RESULTS}

\subsection{Kinematics of Gas and Stars}

\begin{figure*}[t]
\centering
\includegraphics[width=0.9\textwidth]{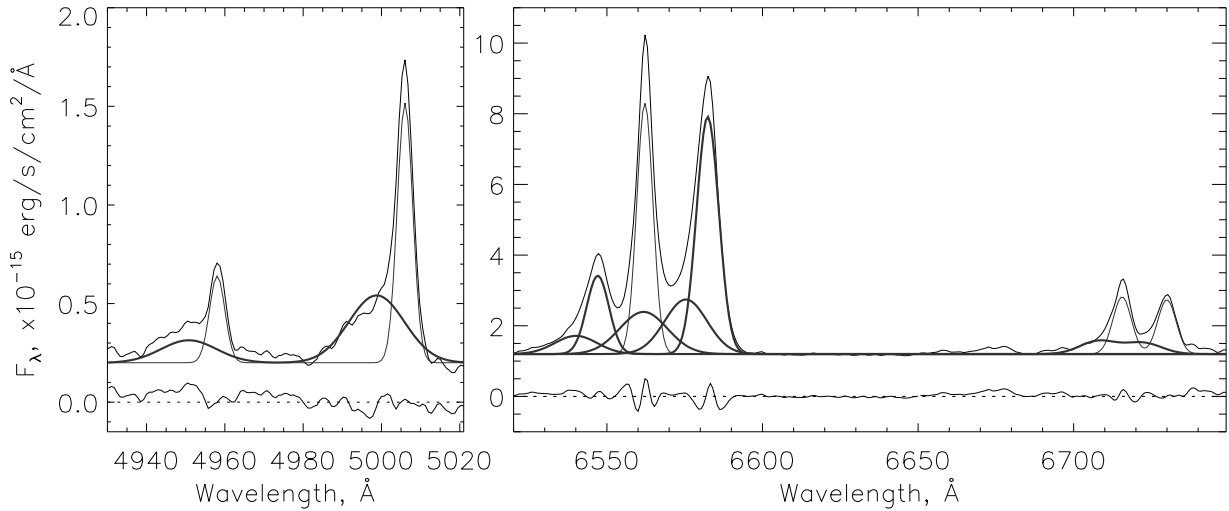}
\caption{Integrated rest--frame spectrum of the region where the asymmetry of
spectral-line profiles is observed according to MPFS data. Left:
emission lines of the [O\,III] doublet. Right: H$\alpha$, [N\,II],
and [S\,II] lines. Also shown is the decomposition of the profiles
into the narrow and broad Gaussian components and the residuals
obtained after subtracting these components. The spectra are
shifted along the vertical axis for convenience.} \label{fig:spec}
\end{figure*}

According to  MPFS data, the distribution of radial velocities of
ionized gas in the \mbox{$r<5\arcsec$--$7\arcsec$} agrees, on the
whole, with the pattern expected for a flat rotating galactic disk
(Fig.~\ref{fig:Halpha}). This conclusion is also confirmed by FPI
observations (Fig.~\ref{fig:IFP}) in the H$\alpha$ line. In this
region the velocity field is quite consistent with H\,I radio
data~\citep{Fer10}.  However, outside the region observed with the
MPFS the velocity field of ionized gas appears significantly
different from circular rotation, especially West of the nucleus.
At the same time, the radial velocities of the outer emission-line
region Northeast of the nucleus (the possible tidal galaxy) agree
with the systemic velocity of the center of  Mkn\,938.

An analysis of the velocity field made by using the
tilted-rings method to determine the following parameters in the
narrow ring at the given distance $r$ from the nucleus: rotation
velocity $V_{\rm rot}$, position angle of the kinematic major axis
$PA_{\rm kin}$, inclination $i$ with respect to the line-of-sight,
and systemic velocity $V_{\rm sys}$. A more detailed description
of the application of this method to FPI and MPFS data can be
found in our earlier papers~\citep{Smirnova2010,Smirnova2018}. We
fixed the inclination at $i=38\degr$, in accordance with
photometric estimates~\citet{Rot04} for  isophotes of the inner
disk in the $K$ band, which is free from dust absorption. We also
fixed $V_{\rm sys}$ in accordance with the mean value averaged
over the field.

Firstly we determined the position of the kinematic center of
rotation based on the assumption that the velocity field is
symmetric. In ionized gas (H$\alpha$\, measurements with  MPFS and
FPI) the center of rotation coincides with the center of continuum
isophotes. However, in the velocity field of stars the kinematic
center is offset North with respect to the photometric center by
about $1\arcsec$.

Fig.~\ref{fig:rc} shows the distribution of kinematic parameters
along the radius. As is evident from the figure, \mbox{$PA_{\rm
kin}\approx-20\degr$} in the inner region, which is close to the
available orientation estimates for the inner disk of the galaxy
in the optical~\citep{Sch07} and near IR~\citep{Rot04}:\linebreak
\mbox{$PA_{\rm phot}\approx-10\degr$}. The small difference
between the position angles can be  explained by radial
mass motions in the circumnuclear spiral structure, which is quite
conspicuous in HST images (see Fig.~\ref{fig:IFP}). At large
distances from the nucleus $PA_{\rm kin}$ in ionized gas begins to
strongly deviate from the photometric major axis, which is due
either to the increase of noncircular (radial) gas motions or to
the warp of the gaseous disk. However, the latter is less likely
because the rotation curve at $r=3\arcsec$--$10\arcsec$ remains
practically flat. The parameters of the rotation of stars at
$r>4\arcsec$ differ appreciably from this of gas rotation: the
kinematic major axis deviates into the opposite direction from the
photometric major axis, and the rotation velocity for the adopted
$i$ decreases abruptly. Note that in the inner regions  $V_{\rm
rot}$ for stars is also almost twice smaller than for ionized gas.
Such a difference is difficult to explain only by asymmetric drift
(higher velocity dispersion) in the stellar disk. Most likely, we
see two dynamic subsystems (the main galaxy and the satellite that
merges with it) in the stellar velocity field. The superposition
of their line-of-sight velocities complicates the observed pattern
and makes it impossible to interpret in terms of the model of the
circular rotation of a flat disk. At the same time, in the
dissipative gaseous system we observe what is already steady
quasicircular motion of gaseous clouds.

\subsection{Ionized Gas Outflow}
\label{sec:outflow} According to MPFS data, the profiles of
emission lines both in the nucleus and within  $r<4\arcsec$ South
of the photometric center show appreciable blue asymmetry. To
investigate it with the best signal-to-noise ratio, we coadded the
spectra in this region after converting to the rest frame in
accordance with the H$\alpha$ velocity field. Fig.~\ref{fig:spec}
shows that the integrated spectrum in each emission line
(H$\alpha$\, and the [N\,II]\,$\lambda$6548,\,6583,
[S\,II]\,$\lambda$6717,\,6731, [O\,III]\,$\lambda$4959,\,5007
doublets) can be described by at least two Gaussian components---a
narrow component with width close to instrumental profile FWHM  of the
spectrograph and another, broader
($FWHM\approx700$--$1000$~km\,s$^{-1}$) component. Note that the
broad component in all forbidden lines is appreciably blueshifted,
the relative radial velocities are equal to about $-370$, $-380$,
and $-480$~km\,s$^{-1}$ in the  [N\,II], [S\,II], and [O\,III]
lines, respectively.

\begin{figure}[]
\centering
\includegraphics[width=\columnwidth]{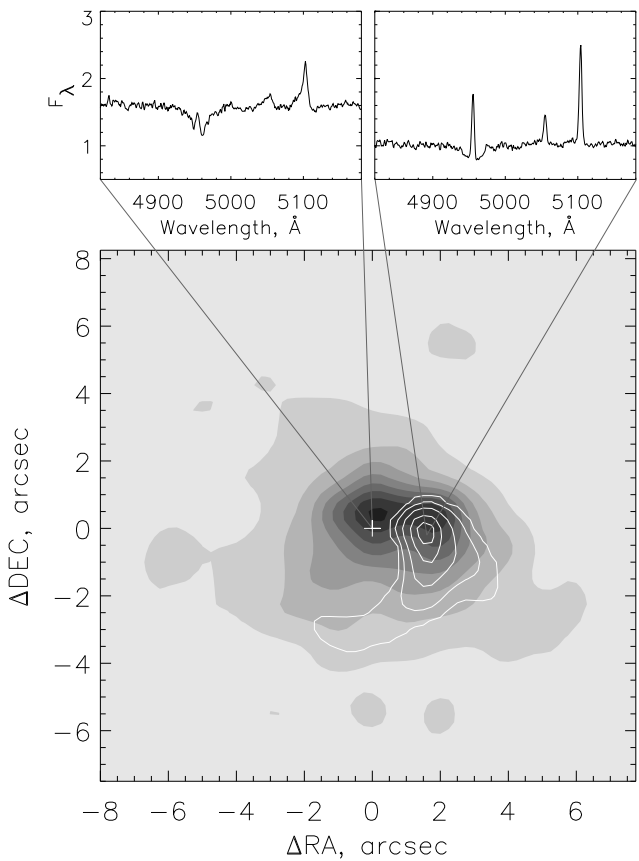}
\caption{H$\beta$ emission isophotes superimposed onto the
[O\,III]\,5007  image with examples of spectra in the selected regions.} \label{fig:Hbeta}
\end{figure}

\label{sec:NaD}

At the same time, the broad   H$\alpha$ component has almost zero
velocity relative to the narrow component. Unfortunately, we could
not perform a similar analysis for the H$\beta$ emission because
of its low brightness and blending by the stellar  H$\beta$
absorption. However, our data allow us to map the distribution of
integrated brightness of  [O\,III] lines and deblended  H$\beta$
emission as shown in Fig.~\ref{fig:Hbeta}. As is evident from the
figure, the brightness distributions in the [O\,III] and H$\beta$
lines differ appreciably. Whereas brightness maximum in [O\,III]
is located at the center of the galaxy, the corresponding peak of
the  H$\beta$\ brightness is shifted Westward by~2\arcsec.
Isophotes of the   H$\beta$\, image have ark-like shape with
centers shifted Southwest by \mbox{1.5--2\arcsec}. A similar
offset is, according to FPI and MPFS data, also observed in the
H$\alpha$\ line. The average radial velocity determined in the
H$\beta$\, absorption is equal to $5890$~km\,s$^{-1}$,which
coincides with the systemic velocity of $5880$~km\,s$^{-1}$. The
narrow  H$\beta$\,  emission line, on the contrary, is shifted by
\mbox{$-200$~km\,s$^{-1}$.}

The observed pattern can be interpreted as wind outflow from the
nucleus with velocities of at least $400$--$500$~km\,s$^{-1}$,
which is often observed in galaxies with powerful starburst. Note
that the galactic wind is dominated by shock
ionization~\citep{Heckman1990,Westmoquette2012}, as is also
evidenced by the shift of the narrow  H$\beta$ line. The presence
of unshifted broad component in the H$\alpha$\, line is indicative
of weak nuclear activity and is associated with the broad
emission-line region  (BLR).

The observed asymmetry can be alternatively explained by the
jet-cloud interac\-tion  \citep[see][]{Smir07}. However, detailed
radio observations with a high spatial resolution revealed no
extended structures that could be interpreted as a
jet~\citep{Fern14}.

\begin{figure*}[]
\centering
\includegraphics[width=\textwidth]{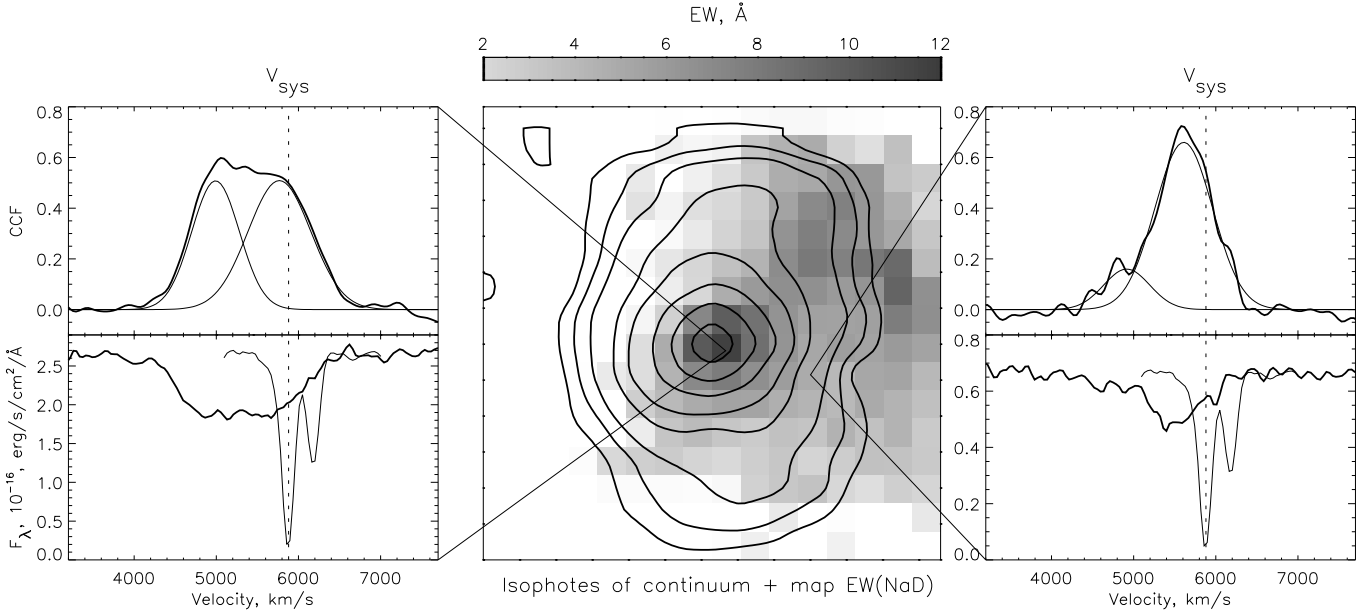}
\caption{Map of the distribution of equivalent widths of  Na\,D line with 5500--6300~\AA\  continuum brightness isophotes superimposed and examples of spectra and their analysis in various regions of the image. The thin line shows an invert spectrum of the airglow Na\,D emission line.
} \label{fig:NaD}
\end{figure*}

\subsection{Galactic Wind in the Extended  Na\,D Line}

\begin{figure}[]
\centering
\includegraphics[width=0.92\columnwidth]{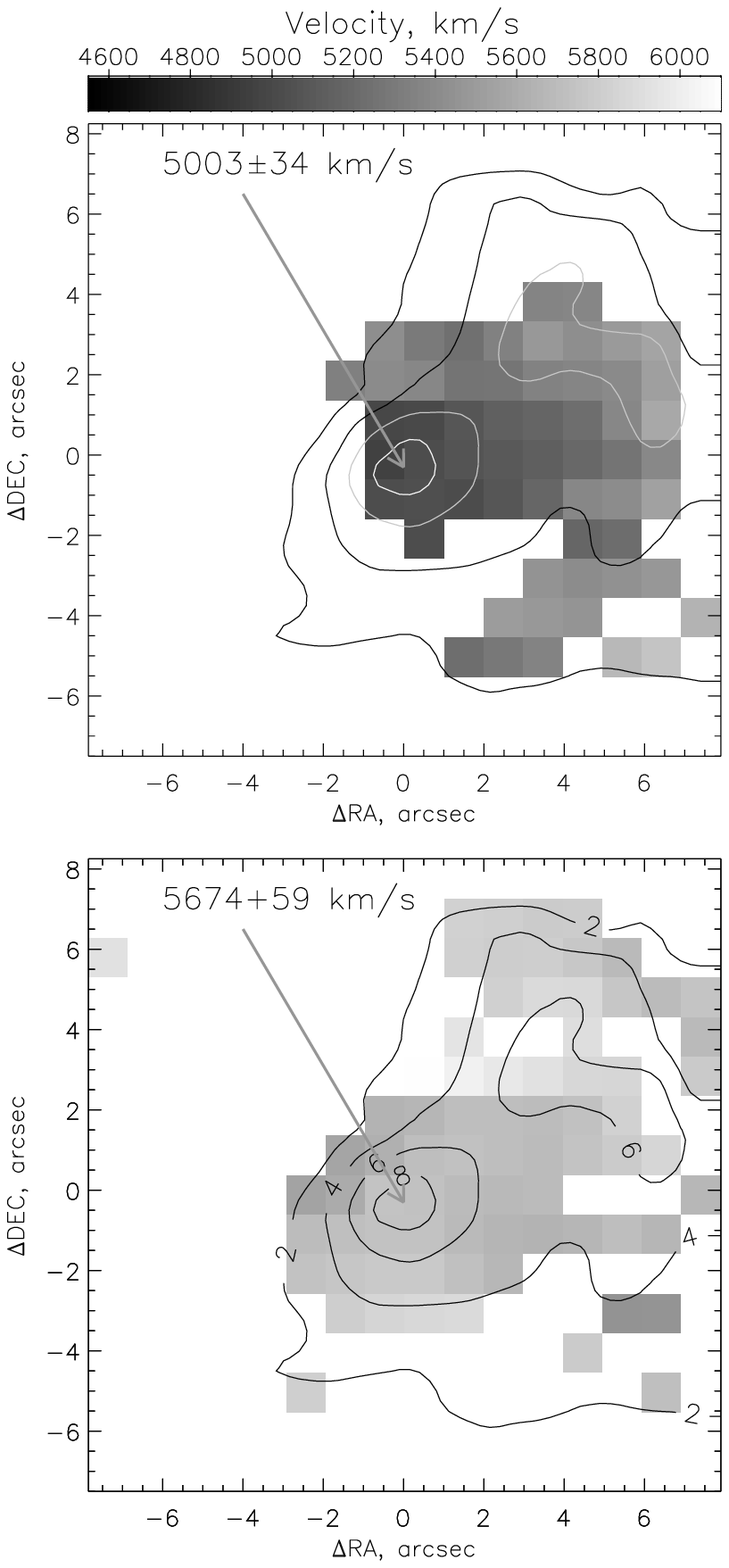}
\caption{Velocity field of the high- (the top panel) and
low-velocity (the bottom panel)  Na\,D components. The isophotes
show the distribution of the equivalent widths of  Na\,D.}
\label{fig:compNaD}
\end{figure}

The spectrum of the central part of the galaxy exhibits a broad
Na\,D, absorption line blueshifted by  $620$~km\,s$^{-1}$ relative
to the systemic velocity of $5880$~km\,s$^{-1}$~\citep{Sch07}.
Such lines, which demonstrate high-velocity outflow from the
center, are observed in many ultrabright infrared active starburst
galaxies~\citep{Rupke2005}.  We used MPFS data to analyze the
variation of observed sodium doublet profiles. Fig.~\ref{fig:NaD}
illustrates the idea of our analysis. The central panel shows the
map of the distribution of equivalent widths of  Na\,D, absorption
lines, which we computed in the wavelength interval
5970--6020~\AA. The left and right panels show the  Na\,D line
profiles and the result of the cross correlation with the Na\,D
lines of the night sky. The velocity scale is printed along the
horizontal axis. Negative night-sky line profiles shifted along
$Z$ are shown next to the profiles of the object lines. The
vertical dashed line indicates the systemic velocity of the
galaxy. The cross-correlation curves are actually the profiles of
the distribution of Na\,D radial velocities at the corresponding
point of the image. A decomposition of these profiles showed that
their shape can be described quite well by two Gaussians of
approximately the same width within
\mbox{$FWHM=600$--$900$~km\,s$^{-1}$}. Fig.~\ref{fig:compNaD}
shows line-of-sight velocity fields for each component. Hereafter we refer
to the highly blueshifted feature the high-velocity component and
the other one, the low-velocity component. Isophotes in the figure
show the distribution of Na\,D equivalent widths. The
high-velocity component at the center has a velocity of
$-800$~km\,s$^{-1}$ relative to the systemic velocity and its
width is of about $700$~km\,s$^{-1}$, the corresponding parameters
of the low-velocity component are~$-200$~km\,s$^{-1}$ and
$900$~km\,s$^{-1}$, respectively.

\section{DISCUSSION}
We observe in  Mkn\,938 the final stage of a merger of two
gas-rich galaxies with dominating star formation in the center.
Our observations revealed many peculiarities of the stellar and
gas kinematics in this galaxy. First, emission-line profiles both
in the nucleus proper and  within $r<4\arcsec$ South of the
photometric center show appreciable blue asymmetry. Whereas in all
forbidden lines the broad component is appreciably blueshifted
(relative radial velocities range from $-370$ to
$-480$~km\,s$^{-1}$ in different lines), the broad component in
the H$\alpha$\ line has almost zero velocity relative to the
systemtic velocity. The presence of the unshifted component in the
H$\alpha$\ line is due to the broad emission-line region (BLR)
around the active nucleus of Mkn\,938.

The observed pattern can be interpreted as wind outflow with
velocities of  \mbox{$400$--$500$~km\,s$^{-1}$} or higher from the
nucleus, as is observed in galaxies with intense star formation.
Note that the galactic wind is dominated by shock
ionization~\citep{Heckman1990,Westmoquette2012}, which is also
evidenced by the shift of the narrow H$\beta$ line. The presence
of unshifted broad  H$\alpha$\, component in the center is
indicative of weak activity of the nucleus and is associated with
the broad emission-line region (BLR).

The line asymmetry is most likely due to an outflow from the
active nucleus. The results of other authors estimating the
velocity of gas motion from the center of Mkn\,938 based on Na\,D
lines to range from~$-620$~km\,s$^{-1}$ to~$-1050$~km\,s$^{-1}$
\citep{Sch07} further confirm this hypothesis. Similar  outflows can
also be observed in starburst galaxies, however  outflow
velocities in Mkn\,938 are too high even for a star-formation rate of
\mbox{$70\pm20~M_\odot\,$yr$^{-1}$}~\citep{Sch07} and can be
explained only the influence of the active nucleus. For example,
by the presence of a jet, which so far has not been found in radio
observations.

High-velocity outflows of neutral gas in the Na\,D absorption are
observed quite often in galactic winds. Such observations are
usually made using the long-slit spectroscopy and therefore the
spatial structure and extension of the outflows remains unclear.
Outflow line-of-sight velocity fields in   Na\,D have so far been
constructed only for a few galactic winds (see, e.g., our earlier
paper about Mkn\,334~\citep{Smirnova2010} or a recent study of the
wind in NGC\,5394~\citep{windN5394}). Note that the spatial size
of the neutral-gas outflow in Mkn\,938 is significantly larger,
and high-velocity gas is observed out to 3~kpc projected distance
from the center. We plan to construct a detailed geometric model
of this outflow in a separate paper including new observational
data.

We  mapped  the distributions of the integrated
brightness of the [O\,III] lines and deblended  H$\beta$\ emission
(see Fig.~\ref{fig:Hbeta}). Whereas the  [O\,III] brightness
maximum coincides with the continuum image of the galaxy nucleus,
the H$\beta$\ brightness maximum is offset~2\arcsec to the West.
Note that the H$\beta$\ emission is shifted by
\mbox{$-200$~km\,s$^{-1}$} relative to the systemic velocity. This
effect can be most likely explained by the gas outflow from the
nucleus with velocities amounting to
\mbox{$400$--$500$~km\,s$^{-1}$}. Gas outflow with a velocity of
about~$400$~km\,s$^{-1}$ was also discovered in  ALMA observations
\citep[see][]{Xu14}.

ALMA observations of Mkn\,938 in the CO\,(6\mbox{--}5) emission
line revealed a dynamic structure, which the authors interpreted
as a rotating circumnuclear disk~\citep{Xu14}. However, this disk
proves to be highly shifted in velocity terms (the velocity of the
disk center is about~$5700$~km\,s$^{-1}$), which is
almost~$200$~km\,s$^{-1}$ less that the systemic velocity. This
shift of the dynamic center can be explained by the fact that what
we actually observe in the CO\,(6\mbox{--}5) line is not a disk
but rather an outflow with velocities on the order
of~$200$~km\,s$^{-1}$, and the galaxy nucleus then corresponds to
velocities of about~$5900$~km\,s$^{-1}$. This systemic velocity
agrees well with numerous observations of other authors
(\mbox{$V_{\rm sys}=5931\pm11$~km\,s$^{-1}$}---RC3/NED,
\mbox{$5881\pm2$~km\,s$^{-1}$}---\citet{Rot06} etc).

\section{CONCLUSIONS}
We performed a detailed study of the active galaxy Mkn\,938 using
methods of panoramic spectroscopy. We investigated
the peculiarities of the structure of Mkn\,938 on different scale
lengths, namely:

\begin{itemize}

\item At large
galactocentric distances ($r>4\arcsec$) the parameters of the
rotation of stars differ appreciably from those of gas rotation.
The stellar velocity field shows at least two dynamic subsystems:
the main galaxy and the nucleus of a companion merging with it,
and therefore the superposition of their line-of-sight velocities
complicates the observed pattern preventing its interpretation in
terms of the model of circular rotation of a flat disk. At the
same time, in the gaseous system we observe steady quasi-circular
motion of gaseous clouds, which can be used to reconstruct a
realistic rotation curve.

\item We mapped for the first  time in Mkn~938 the galactic wind in the Na\,D
absorption line. A decomposition of the profiles of this line
showed that their shape can be described quite well by two
Gaussian of about the same width within
\mbox{$FWHM=600$--$900$~km\,s$^{-1}$}. We constructed the velocity
fields for both, the high- and low-velocity  components. The
high-velocity component in the center shows a velocity of
$-800$~km\,s$^{-1}$ relative to the systemic velocity and a width
of about~$700$~km\,s$^{-1}$. The corresponding parameters for the
low-velocity component are \mbox{$-200$~km\,s$^{-1}$} and
$900$~km\,s$^{-1}$, respectively.

\item In the innermost region of the galaxy within $r<4\arcsec$
South of the photometric center a second component blueshifted by
up to~\mbox{$-500$~km\,s$^{-1}$} is also observed in the emission
lines of ionized gas. Furthermore, the brightness center in Balmer
lines is appreciably (almost by~$2\arcsec$ or about~$0.8$~kpc)
offset relative to center of the [O\,III] and continuum isophotes.
All this together can be interpreted either as the ionized
component of galactic wind mapped in  Na\,D, or as the effect of
the jet emerging from the active nucleus on the interstellar
medium. Unfortunately, no jet has so far been detected in radio
observations of   Mkn\,938.

\end{itemize}

\label{S:con}

\begin{acknowledgements}
We are grateful to the anonymous referee for the comments, which
helped to improve the paper. This research has made use of the
NASA/IPAC Extragalactic Database (NED), which is operated by the
Jet Propulsion Laboratory, California Institute of Technology,
under contract with the National Aeronautics and Space
Administration. In this paper we present an image from the  
 Hubble Legacy Archive  based on observations made with the
NASA/ESA Hubble Space Telescope, and obtained from the Hubble
Legacy Archive, which is a collaboration between the Space
Telescope Science Institute (STScI/NASA), the Space Telescope
European Coordinating Facility (ST-ECF/ESA) and the Canadian
Astronomy Data Centre (CADC).
\end{acknowledgements}

\section*{FUNDING}
This work was supported by the Russian Science Foundation (project
No.~17-12-01335 ``Ionized gas in galaxy disks and beyond the
optical radius'').  Observations with the SAO RAS telescopes are
supported by the Ministry of Science and Higher Education of the
Russian Federation (including agreement No. 05.619.21.0016,
project ID RFMEFI61919X0016).


\end{document}